\documentclass[twocolumn]{article}
\usepackage[top=2cm,bottom=2cm,left=2cm,right=2cm]{geometry}
\usepackage{framed,multirow}

\usepackage{amssymb}
\usepackage{latexsym}

\usepackage{url}
\usepackage{xcolor}
\definecolor{newcolor}{rgb}{.8,.349,.1}
\usepackage[allbordercolors=white]{hyperref}

\usepackage[ruled,vlined]{algorithm2e}

\usepackage{amsmath}
\usepackage{tikz}
\usepackage{booktabs}
\usepackage{natbib}

\usepackage{array}
\newcolumntype{L}[1]{>{\raggedright\let\newline\\\arraybackslash\hspace{0pt}}m{#1}}
\newcolumntype{C}[1]{>{\centering\let\newline\\\arraybackslash\hspace{0pt}}m{#1}}
\newcolumntype{R}[1]{>{\raggedleft\let\newline\\\arraybackslash\hspace{0pt}}m{#1}}

\newcommand{\mc}[3]{
  \multicolumn{#1}{#2}{#3}
}

\newcommand{\mr}[2]{
  \multirow{#1}{*}{#2}
}

\begin{document}

\title{DCDistance: A Supervised Text Document Feature extraction based on class labels}

\author{Charles Henrique Porto Ferreira\\
Debora Maria Rossi de Medeiros\\
Fabricio Olivetti de Franca} 

\maketitle
  



\begin{abstract}
Text Mining is a field that aims at extracting information from textual data. One of the challenges of such field of study comes from the pre-processing stage in which a vector (and structured) representation should be extracted from unstructured data. The common extraction creates large and sparse vectors representing the importance of each term to a document. As such, this usually leads to the curse-of-dimensionality that plagues most machine learning algorithms. To cope with this issue, in this paper we propose a new supervised feature extraction and reduction algorithm, named DCDistance, that creates features based on the distance between a document to a representative of each class label. As such, the proposed technique can reduce the features set in more than $99\%$ of the original set. Additionally, this algorithm was also capable of improving the classification accuracy over a set of benchmark datasets when compared to traditional and state-of-the-art features selection algorithms.
\end{abstract}
\section{Introduction}

Many interesting datasets are acquired from textual information such as products review, social network posts, text articles, e-mails, etc. For this reason, Text Mining is an extensively researched topic in Computer Science.

The main goal of Text Mining is to automatically extract information and knowledge from text documents~\citep{book:witten2016dataMining}. Some common tasks associated with Text Mining are classification algorithms (i.e., topic extraction, sentiment analysis, subject)~\citep{KUMAR2016128}, data clustering (i.e., plagiarism detection, summarization)~\cite{WANG201726}, regression analysis (i.e., recommender systems)~\cite{AGUILAR2017147}. Each of these tasks has its own particularities, specific challenges and applications.

There are many Machine Learning algorithms available to perform these tasks, many of them expect that the input is represented as a vector. As such, an important step for Text Mining is the feature extraction and, as we will highlight next, also the feature selection. 

A common approach is the document vectorization as a bag-of-features~\citep{manning2008introduction} in which a document $i$ is first tokenized, usually by words, and a vector $t_i$ is created with each element corresponding to a token. The value of an element $t_{i,j}$ is defined by:

\begin{equation*}
    t_{i,j} = \left\{\begin{matrix}
 w_{i,j} & \text{ if the token } j \text{is present in document } i\\
 0 & \text{ otherwise}
\end{matrix}\right.,
\end{equation*}

\noindent where $w_{i,j}$ is a weighting of the token $j$ in document $i$.

As one may notice, this can potentially lead to very sparse and high-dimensional vectors due to the diversity of tokens in a corpus.

This can be a problem to many Machine Learning algorithms that suffer from the curse-of-the-dimensionality~\citep{book:1643328} and can also increase the computational costs unnecessarily.

As such, different algorithms were proposed in the literature in order to reduce the dimensionality of the features set while either improving the accuracy of the task or minimizing the impact on the accuracy value.



Specifically for the classification task, some of these algorithms use a ranking method to select the most suitable features, also known as filter methods~\citep{article:Rehman20153670,Stanczyk:2014:FSD:2728712}. Another group of algorithms use a classification algorithm to verify efficiency of a set of features. This method is called wrapper~\citep{Stanczyk:2014:FSD:2728712}. 

One disadvantage of the current pre-processing algorithms is the amount of time required to perform the computation of the feature importance or transformation. It is noticed that a significant amount of time of creating a Machine Learning model for Text Mining corresponds to the pre-processing stage~\citep{MUNKOVA20131198}. Another common disadvantage is related to the number of generated or selected features, that is usually a parameter of the algorithm, thus requiring a fine-tuning to find a compromise between dimensionality and accuracy.

In order to alleviate these problems, in this paper, we propose the DCDistance algorithm. This algorithm extract features solely based on the distance between text documents and representative points of each label. 

Simply put, for a dataset with $k$ distinct labels, this algorithm will create a representation for each document as a $k$-dimensional vector $d_i$ with each element $d_{i,j}$ representing the distance of document $i$ to class $j$.

As a result, DCDistance is capable of reducing the dimension of Bag-of-Words representation in more than $99\%$ of the original size. Also, since this representation contains supervised information regarding the class labels, it often helped to improve the classification accuracy for some of the tested datasets when compared to the application of the original features and some other feature reduction techniques.

The remainder of this paper has the following structure. Section \ref{sec:feature_selection_and_extraction} will present some core concepts together with a selection of the most recent work related to this paper. Section \ref{sec:document_class_distance} will present the proposed algorithm in details. Section \ref{sec:experiments} describes the experiments performed to assess the performance of our proposal and the results are presented in Section~\ref{sec:results}. Finally, some final remarks and summary of the obtained results are given in Section~\ref{sec:conclusion_and_future_work}.

\section{Feature Selection and Extraction}\label{sec:feature_selection_and_extraction}

The extraction and further selection of the features play an important role during pre-processing of text documents since they can reduce noise and sparsity which often improves the model accuracy of the task at hand.

Feature selection techniques select a subset of the original features set, choosing the $k$ most relevant features according to a given criteria. For example, two common criteria are the Information Gain of a feature~\citep{Forman20031289} and the Chi-Squared~\citep{manning2008introduction}.

Information Gain measures the amount of information that a feature can provide when used to discriminate between classes. Similarly, the Chi-Squared evaluates the degree of dependency between a feature and a label. The greater the Chi square's score, the greater the dependence of the feature and that specific label.  

Feature extraction techniques, on the other hand, are applied to the original set of features in order to generate a new and more informative set. The objective of these techniques is that the transformed feature space becomes easier to separate by the traditional classification algorithms. 
\subsection{Recent Work}

\cite{UYSAL201682} proposed a feature selection approach named Improved Global Feature Selection(IGFSS) which combine Global Feature Selection(GFS) techniques~\citep{Guyon:2003:IVF:944919.944968} with Local Feature Selection(LFS) techniques~\citep{TasCi:2013:CTF:2479984.2480039}. A GFS algorithm is a filter selection approach that ranks the features w.r.t. the entire dataset and then select the top $k$ features. The Information Gain and Chi-Squared, explained above, are both examples of GFS.

The LFS algorithms rank the features w.r.t. the classes of the dataset, thus, each feature receive a rank based on each class. With this information, it is possible to assign a label to each feature. Odds Ratio~\citep{manning2008introduction,Forman20031289} and Correlation Coefficient are examples of LFS.

The authors argued that the classical GFS techniques not select enough features that belong to each class of the dataset, so they proposed to select an equal number of features from each class using the information acquired with the GFS and LFS approaches. This technique was compared with some classical GFS techniques such as IG, DFS~\citep{Uysal:2012:NPF:2396898.2397010} and Gini Index~\citep{Shang:2007:NFS:1223505.1223633} and overcame them in all of the four tested benchmarks.

\cite{AGNIHOTRI2017268} proposed a feature selection algorithm similar to \cite{UYSAL201682} using GFS and LFS to select features distributed among the classes. However, they argued that an equal number of features per class may not be reasonable because of the imbalance found in class distributions in many datasets. Instead, they select a number of features proportional to the number of samples belonging to each class. This technique was compared with IG, Mutual Information, Gain Ratio, DFS and the state-of-the-art IGFSS. The results showed significant improvements regarding F-micro and F-macro classification.

\section{Document-Class Distance}\label{sec:document_class_distance}

The main contribution of this paper is the proposal of a feature extraction algorithm that compacts the dimensionality of the dataset while maintaining discriminative power for classification tasks.

Basically, a vector of the same size of the number of distinct labels represents each text document. So, the element $d_{i,j}$ represents the distance from the $i$-th text document to a representative vector of the $j$-th class.

The algorithm, named DCDistance, is depicted in Alg.~\ref{alg:dcdistance} and explained further in the following paragraphs.

\begin{algorithm}[ht!]
  \SetKwFunction{vectorize}{vectorize}\SetKwFunction{sum}{sum}\SetKwFunction{filter}{filter}\SetKwFunction{dist}{dist}
  \SetKwFunction{stoplistRemoval}{stoplistRemoval}
  \SetKwFunction{porterStemming}{porterStemming}
  \SetKwInOut{Input}{input}\SetKwInOut{Output}{output}
  \Input{train and test text documents $D_{train}$ and $D_{test}$, set of labels $Y_{train}$ and number of distinct labels $k$.}
\Output{vectorial representation $V'_{train}, V'_{test}$}
\BlankLine

\stoplistRemoval{$D_{train}$}\;
\porterStemming{$D_{train}$}\;
$V_{train} \leftarrow$ \vectorize{$D_{train}$}\;
$V_{test} \leftarrow$ \vectorize{$D_{test}$}\;
\For{$i = 1 \dots k$}{
    $Vd[i] \leftarrow$ \sum{\filter{$y==i, (V_{train},Y_{train})$}}\;
}

\For{$i = 1 \dots |V_{train}|$}{
   $V'_{train}[i] \leftarrow$ [\dist{$V_{train}[i]$, Vd$[j]$} $\mid j \leftarrow 1 \dots k$]\;
}
\For{$i = 1 \dots |V_{test}|$}{
   $V'_{test}[i] \leftarrow$ [\dist{$V_{test}[i]$, Vd$[j]$} $\mid j \leftarrow 1 \dots k$]\;
}
\Return $V'_{train}, V'_{test}$\;
\caption{DCDistance algorithm.}
\label{alg:dcdistance}
\end{algorithm}

The algorithm starts with the application of any vectorization algorithm on the corpus data. For example, a Bag-of-Words with TF-IDF weighting. After that, the vectors of the documents corresponding to each label are summed up generating a representative vector of this particular class. Finally, the new vector representation for each document is created by calculating the distance between this document and each representative vector.

Notice that the $dist$ function can be any distance applicable into a multi-dimensional numerical vector. The summing of the representative vectors works as an aggregation of the information contained in each document vector. Also, it should be noticed that the representative vectors are built without the test data information.

Regarding the computational complexity, after the vectorization step, the algorithm performs $O(n \cdot d)$  + $O(p \cdot d)$ operations, with $n$ as the number of documents of the training set, $p$ as the number of documents from the test set and $d$ the dimension of the feature vector generated by the vectorization. The new vector space is created with $O(m \cdot k \cdot d)$ operations, with $m = n+p$ the total number of documents and $k$ the number of distinct class labels. As such, the overall complexity is proportional to $O(m \cdot k \cdot d)$.

\section{Experiments}\label{sec:experiments}

\begin{figure}[htb]
\resizebox{\columnwidth}{!}{
  \includegraphics{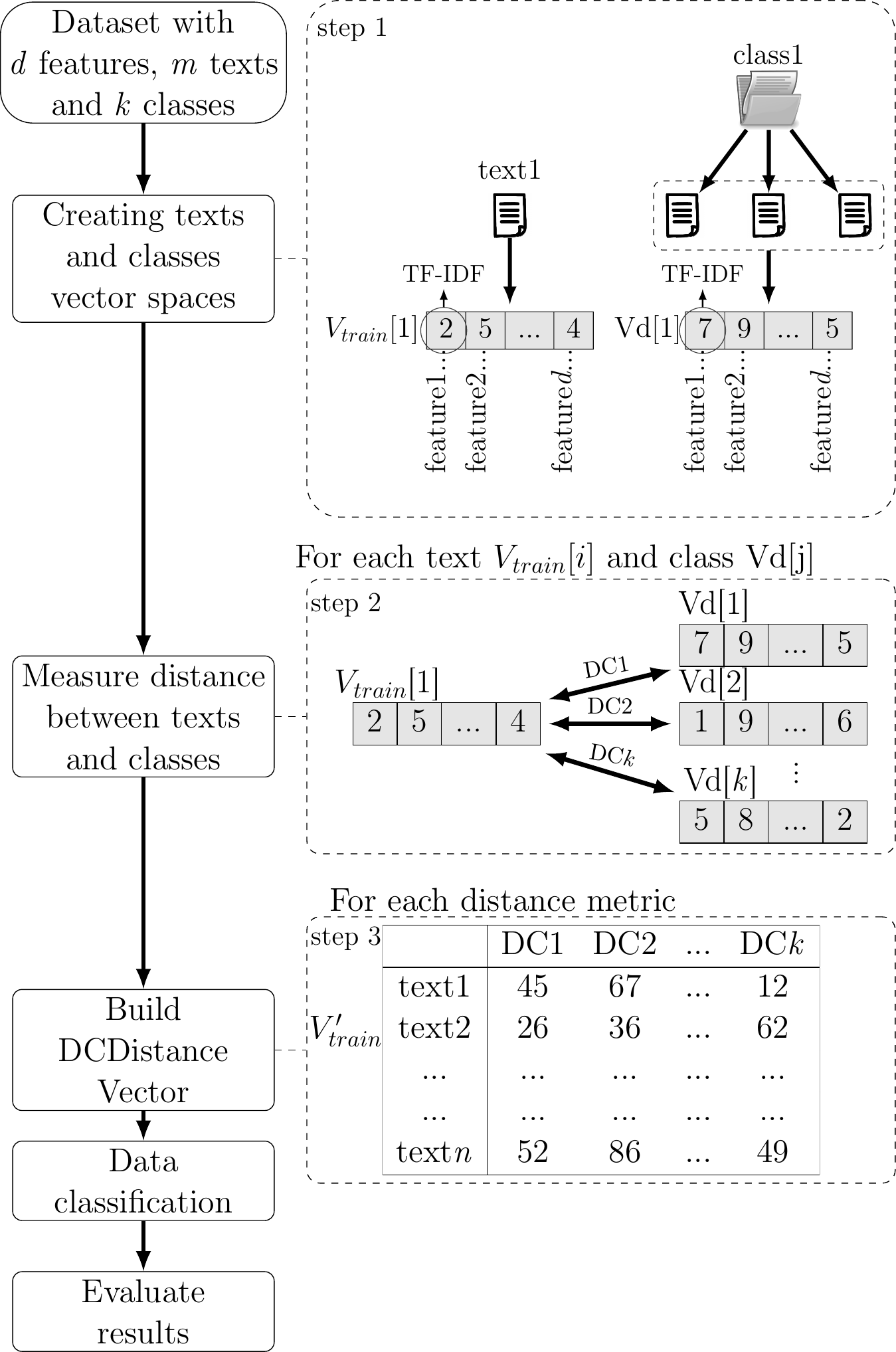}}
  \caption{Flowchart of the proposed technique. Step 2 and 3 are exemplified with $V_{train}$ to generate $V'_{train}$, but they must be done, again, with $V_{test}$ to generate $V'_{test}$, as showed on \autoref{alg:dcdistance}}
  \label{fig:fluxograma}
\end{figure}

In order to assess whether this representation retains enough information for classification tasks, we have devised an experimentation pipeline as depicted in Fig.~\ref{fig:fluxograma}. In short, we first apply some common pre-processing steps to the text documents such as English stopword removal and Porter Stemming Algorithm~\citep{article:porter1980algorithm}. After that, we tokenized each document and generated a vector representation weighted by TF-IDF such that every position in the vector represents a token from the corpus, and the value of the $i$-th token to the $j$-th document is calculated by:

\begin{equation}
TF-IDF(t_i,d_j) = \\ \left\{\begin{matrix}
f(t_i,d_j) \cdot \log{\frac{N}{|\{d \in D : t_i \in d|\}}}, & \text{if } t_i \in d_j  \\
0, & \text{otherwise}
\end{matrix}\right.,
\end{equation}

\noindent where $f(t,d)$ measures the frequency of a term $t$ in document $d$ and $D$ is the set of all documents.

After that, we have performed a $10$-fold cross-validation in the vectorized dataset and for each combination of train and test we have applied the DCDistance feature extraction algorithm and the others baselines feature selection algorithm as described on algorithm \ref{alg:featureSelection}, thus generating $10$ different train and test transformed data. 

For each one of the training data, we have applied the classifications algorithms Support Vector Machine (SVM)~\citep{weiss2015fundamentals}, $k$-Nearest Neighbors (kNN)~\citep{weiss2015fundamentals} and Random Forest (RF)~\citep{GENUER201728}, all fitted on the training set and evaluated on the test set. 

To measure the performance, we will present the mean classification accuracy, micro-F1 and macro-F1 results on the test data. We compared the results with the TF-IDF vectorization applied to four feature selection algorithms: Information Gain and Chi-squared, IGFSS and VGFSS.

\begin{algorithm}[ht!]
  \SetKwFunction{vectorize}{vectorize}\SetKwFunction{sum}{sum}\SetKwFunction{filter}{filter}
  \SetKwFunction{stoplistRemoval}{stoplistRemoval}
  \SetKwFunction{porterStemming}{porterStemming}
  \SetKwFunction{applyFeatureSelectTechnique}{applyFeatureSelectTechnique}
  \SetKwFunction{selectFeatures}{selectFeatures}
  \SetKwInOut{Input}{input}\SetKwInOut{Output}{output}
  \Input{train and test text documents $D_{train}$ and $D_{test}$}
\BlankLine

\stoplistRemoval{$D_{train}$}\;
\porterStemming{$D_{train}$}\;
$V_{train} \leftarrow$ \vectorize{$D_{train}$}\;
$V_{test} \leftarrow$ \vectorize{$D_{test}$}\;
\applyFeatureSelectTechnique{$V_{train}$}\;
\selectFeatures{$V_{train}$}\;
\selectFeatures{$V_{test}$}\;
\caption{Feature selection process.}
\label{alg:featureSelection}
\end{algorithm}

Since these feature selection algorithms require the number of features to be selected as a parameter, we have tested four different input parameters: i) the same number of features created by DCDistance; ii) $20\%$ of the number of features of the vectorized representation; iii) $40\%$ of the number of features of the vectorized representation; iv) $60\%$ of the number of features of the vectorized representation. In the results tables and plots they were labeled as IG1/Chi1/IGFSS1/VGFSS1, IG2/Chi2/IGFSS2/VGFSS2, IG3/Chi3/IGFSS3/VGFSS3, and IG4/Chi4/IGFSS4/VGFSS4, respectively. We parameterized IGFSS using DFS as GFS and Odds Ratio as LFS. VGFSS was parameterized with DFS as GFS and max of IG, GI, DFS and GR as LFS. The tests were performed on four different benchmark datasets from the literature, with their features summarized in Tab.~\ref{tab:datasets}:

\begin{itemize}

\item \textbf{Reuters-21578:} a well know dataset used in text mining applications (\cite{article:Rehman20153670}). We are using the Apte' split of this dataset provided by \cite{site:corpus_reuter_21578}. This division chooses the ten classes with the largest number of texts.

\item \textbf{20Newsgroup:} a corpus containing $7,532$ documents and $65,981$ distinct words extracted from $20$ different newsgroups, with each one representing a different class. This dataset version is available on \cite{site:20newsgroup} web page. We worked only with "by date" version and used the test set as the whole dataset, i.e., divided the test set into test and training set.

\item \textbf{SCY-Genes:} a dataset collected by~\cite{med04,med05} with the abstracts of scientific papers about the genes of \emph{Saccharomyces Cerevisiae}. The documents are classified according to the type of gene studied.

\item \textbf{SCY-Cluster:} the same as the previous corpus but with documents classified by the studied biological function~\citep{med04,med05}.

\end{itemize}

\begin{table}[htb]
  \caption{Characteristics of the benchmark datasets used during the experiments. Words pp. are the number of words after pre-processing stage and Desv. pad is the standard  deviation  of the number of texts within each class}
  \label{tab:datasets}
  \centering
  \resizebox{\columnwidth}{!}{
  \begin{tabular}{llllll}
    \toprule
      \textbf{Datasets}     & \textbf{Classes}  & \textbf{Docs}   & \textbf{Words}  & \textbf{Words pp.} & \textbf{Desv.pad}    \\
      \midrule  
      Reuters-21578 &   10     & 2787    & 13005  & 9041      & 325.37     \\
      20NewsGroup   &   20     & 7532    & 65981  & 50030     & 37.75      \\
      SCY-Genes     &   7      & 1114    & 9128   & 5963      & 56.21      \\
      SCY-Cluster   &   7      & 1655    & 11273  & 7630      & 158.81     \\
    \bottomrule
  \end{tabular}
  }
\end{table}

Regarding the hyper-parameters of the classifiers, SVM parameters were chosen after a MultiSearch algorithm\footnote{\url{https://github.com/fracpete/multisearch-weka-package}} performed on the training set by choosing the kernel between Polynomial and RBF, the degree for the polynomial kernel values ranging from 1 to 5 and for the RBF kernel, the gamma value raging from -4 to 1, and the C parameter tested within fixed value 1. For the k-NN we have tested the parameter k ranging from $1$ to $15$.

The best parameters obtained for the training data was achieved with polynomial kernel with degree $5$ for the Euclidean distance, RBF kernel with gamma set to $10$ for the cosine distance and polynomial kernel with degree $1$ for the IG, CHI, IGFSS and VGFSS, in all cases $C = 1$. The $k$-NN algorithm achieved the best results with $k = 5$ and Euclidean distance. 

\section{Results}\label{sec:results}

\begin{figure*}[tb]
  \centering
  \resizebox{\textwidth}{!}{
  \includegraphics[]{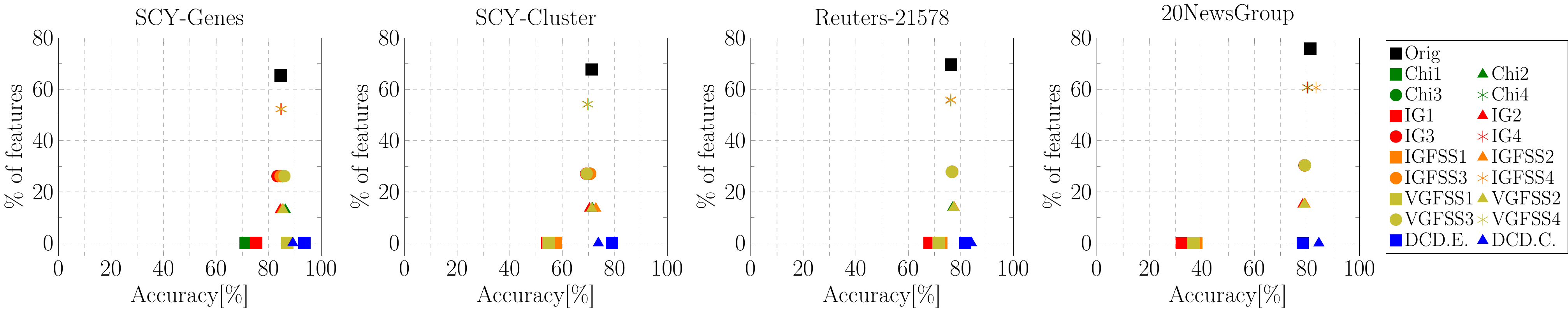}
  }
  \caption{Results achieved with the DCDistance technique applied with the SVM classifier. The ``Accuracy'' axis represents the accuracy of the classification and  the ``Features'' axis represents the resulting percentage of features.}
  \label{fig:graphs_DCD_SVM}
\end{figure*}

\begin{figure*}[tb]
  \centering
  \resizebox{\textwidth}{!}{
  \includegraphics[]{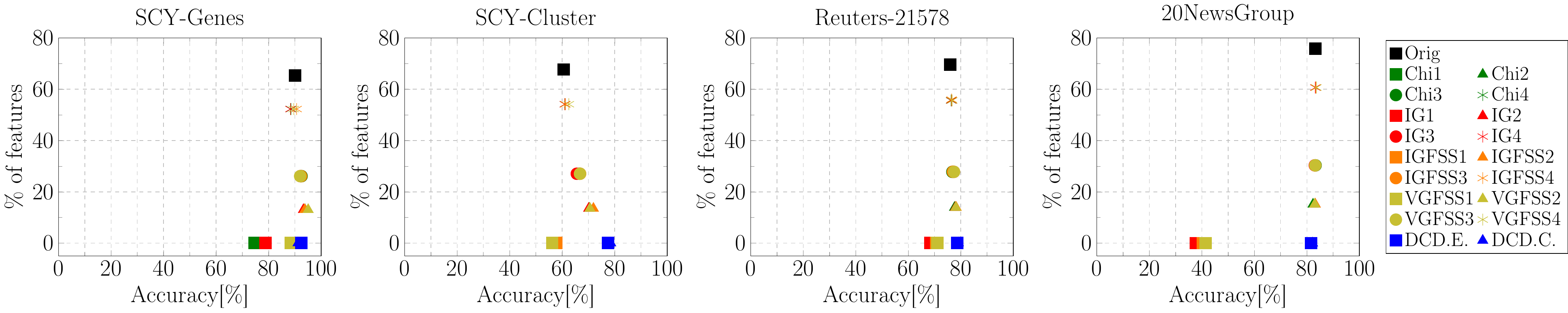}
  }
  \caption{Results achieved with the DCDistance technique applied with the Random Forest classifier. The ``Accuracy'' axis represents the accuracy of the classification and  the ``Features'' axis represents the resulting percentage of features.}
  \label{fig:graphs_DCD_RF}
\end{figure*}

\begin{figure*}[tb]
  \centering
  \resizebox{\textwidth}{!}{
  \includegraphics[]{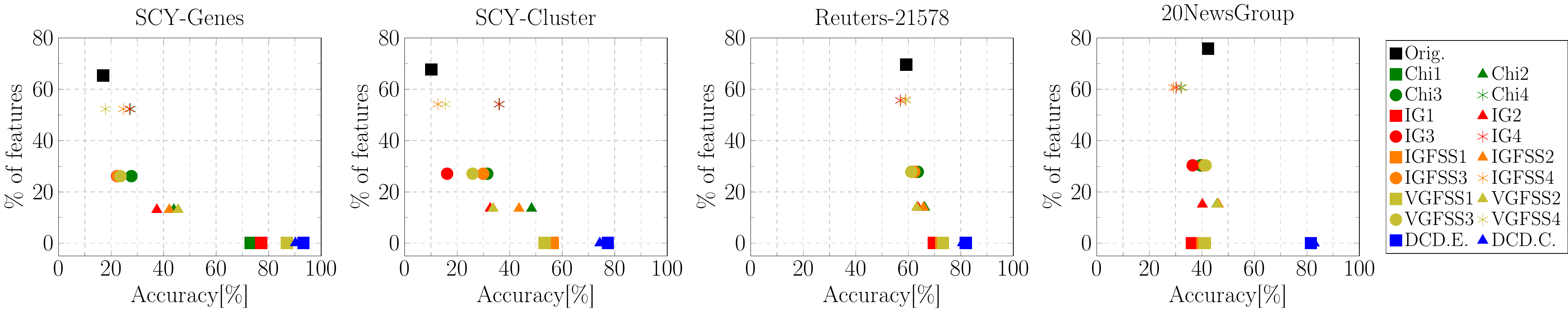}
  }
  \caption{Results achieved with the DCDistance technique applied with the KNN classifier. The ``Accuracy'' axis represents the accuracy of the classification and  the ``Features'' axis represents the resulting percentage of features.}
  \label{fig:graphs_DCD_KNN}
\end{figure*}

\begin{table}[tb]
  \caption{Experiments results with Reuter Dataset}
  \label{tab:results_reuter}
  \centering
  \resizebox{\columnwidth}{!}{
	\begin{tabular}{lccccc}
	\toprule
	\textbf{Tech}				 & \textbf{SVM}			& \textbf{KNN} 	& \textbf{RF} 	& \textbf{Feat.} 	& \textbf{Red(\%)} \\
	\midrule
	Orig    & 76.21$\pm$2.18     & 59.17$\pm$3.95     & 75.96$\pm$2.85      & 9041   			& 34.67      			\\ \midrule
	Chi4    & 76.17$\pm$2.26     & 57.02$\pm$3.50     & 76.32$\pm$2.59      & \mr{4}{7233}   & \mr{4}{47.73}            \\
	IG4     & 76.17$\pm$2.26     & 57.02$\pm$3.40     & 76.46$\pm$2.53      & 		  			&       				\\ 
	IGFSS4  & 76.17$\pm$2.31     & 58.88$\pm$3.28     & 76.50$\pm$2.75      & 		  			&       				\\
	VGFSS4  & 76.25$\pm$2.31     & 59.13$\pm$3.82     & 76.75$\pm$2.32      & 		  			&       				\\ \midrule
	Chi3    & 76.60$\pm$2.48     & 63.62$\pm$2.19     & 76.78$\pm$3.21      & \mr{4}{3616}   & \mr{4}{73.87}            \\
	IG3     & 76.68$\pm$2.12     & 61.36$\pm$3.08     & 77.39$\pm$3.32      & 		   			&       				\\
	IGFSS3  & 76.75$\pm$2.39     & 62.29$\pm$2.49     & 77.00$\pm$2.95      & 		   			&       				\\
	VGFSS3  & 76.60$\pm$2.18     & 61.14$\pm$2.83     & 77.43$\pm$2.96      & 		   			&       				\\ \midrule
	Chi2    & 76.93$\pm$2.52     & 66.24$\pm$2.78     & 77.61$\pm$2.87      & \mr{4}{1808}   & \mr{4}{86.94}            \\
	IG2     & 77.36$\pm$2.40     & 63.72$\pm$2.19     & 77.97$\pm$2.82      & 		  			&       				\\
	IGFSS2  & 77.50$\pm$2.10     & 65.81$\pm$1.81     & 78.26$\pm$2.24      & 		  			&       				\\
	VGFSS2  & 77.29$\pm$1.96     & 63.22$\pm$2.41     & 78.11$\pm$2.45      & 		  			&       				\\ \midrule
	IG1     & 68.03$\pm$2.15     & 69.79$\pm$2.55     & 68.42$\pm$2.65      & \mr{6}{10}     & \mr{6}{99.92}            \\
	Chi1    & 72.05$\pm$3.04     & 71.15$\pm$3.21     & 69.39$\pm$3.06      & 	    			&       				\\ 
	IGFSS1  & 72.59$\pm$1.24     & 72.55$\pm$2.36     & 70.65$\pm$2.06      & 	    			&       				\\
	VGFSS1  & 71.47$\pm$1.95     & 73.23$\pm$2.21     & 71.11$\pm$2.31      & 	    			&       				\\ 
	DCD. E. & 81.67$\pm$3.05     & \bf{81.88$\pm$1.88}& 78.61$\pm$1.96      & 	    			&  	  					\\
	DCD. C. & \bf{84.07$\pm$2.48}& 80.62$\pm$1.76     & \bf{79.12$\pm$1.79} & 					&       				\\
	\bottomrule
	\end{tabular}
	}
\end{table}
\begin{table}[tb]
  \caption{Experiments results with News Dataset}
  \label{tab:results_news}
  \centering
  \resizebox{\columnwidth}{!}{
	\begin{tabular}{lccccc}
	\toprule
	\textbf{Tech}				 & \textbf{SVM}			& \textbf{KNN} 	& \textbf{RF} 	& \textbf{Feat.} 	& \textbf{Red(\%)} \\
	\midrule
	Orig    & 81.33$\pm$1.59     & 42.39$\pm$2.30     & 83.19$\pm$0.86     & 50030  		   & 24.17      		\\ \midrule
	Chi4    & 80.38$\pm$1.32     & 32.05$\pm$2.84     & 83.48$\pm$1.20     & \mr{4}{40024}  & \mr{4}{39.34}   \\
	IG4     & 80.20$\pm$1.09     & 30.25$\pm$3.28     & 83.21$\pm$1.03     &       									\\ 
	IGFSS4  & 83.47$\pm$1.41     & 29.13$\pm$3.81     & 83.47$\pm$1.41     &       									\\
	VGFSS4  & 80.01$\pm$1.39     & 32.44$\pm$6.63     & \bf{83.58$\pm$1.14}&       									\\ \midrule
	Chi3    & 79.25$\pm$1.74     & 39.56$\pm$1.03     & 83.31$\pm$1.32     & \mr{4}{20012}  & \mr{4}{69.67}   \\
	IG3     & 79.09$\pm$1.81     & 36.50$\pm$2.25     & 83.01$\pm$1.59     &       									\\
	IGFSS3  & 79.22$\pm$1.66     & 40.87$\pm$2.11     & 83.18$\pm$0.93     &       									\\
	VGFSS3  & 79.22$\pm$1.80     & 41.41$\pm$1.23     & 83.15$\pm$1.17     &       									\\ \midrule
	Chi2    & 79.09$\pm$1.37     & 45.75$\pm$1.79     & 82.33$\pm$1.09     & \mr{4}{10006}  & \mr{4}{84.83}   \\
	IG2     & 78.47$\pm$1.29     & 40.19$\pm$1.96     & 83.18$\pm$1.14     &      									\\
	IGFSS2  & 79.30$\pm$1.27     & 46.28$\pm$1.74     & 83.19$\pm$0.84     &      									\\
	VGFSS2  & 78.98$\pm$1.74     & 45.86$\pm$1.25     & 82.97$\pm$1.18     &      									\\  \midrule
	Chi1    & 32.38$\pm$1.81     & 38.56$\pm$1.79     & 38.21$\pm$2.15     & \mr{6}{20}     & \mr{6}{99.96}   \\
	IG1     & 32.43$\pm$1.57     & 36.17$\pm$1.86     & 37.75$\pm$2.12     &       									\\
	IGFSS1  & 37.87$\pm$1.67     & 40.29$\pm$1.49     & 40.32$\pm$1.21     &       									\\
	VGFSS1  & 36.78$\pm$2.56     & 41.16$\pm$1.92     & 41.38$\pm$1.69     &       									\\ 
	DCD. E. & 78.41$\pm$1.73     & 81.51$\pm$1.53     & 81.55$\pm$0.86     &  	   									\\ 
	DCD. C. & \bf{84.56$\pm$1.65}& \bf{82.85$\pm$1.41}& 82.16$\pm$1.41     &       									\\       
	\bottomrule
	\end{tabular}
	}
\end{table}
\begin{table}[tb]
  \caption{Classification accuracy comparison obtained on Genes dataset with three different classifiers.}
  \label{tab:results_genes}
  \centering
  \resizebox{\columnwidth}{!}{
	\begin{tabular}{lccccc}
	\toprule
	\textbf{Tech}				 & \textbf{SVM}			& \textbf{KNN} 	& \textbf{RF} 	& \textbf{Feat.} 	& \textbf{Red(\%)} \\
	\midrule
	Orig    & 84.65$\pm$3.01     & 16.97$\pm$3.63     & 90.03$\pm$1.87     & 5963  			& 34.67     		\\ \midrule
	Chi4    & 84.82$\pm$3.52     & 27.37$\pm$4.88     & 88.51$\pm$1.96     & \mr{4}{4771}  	& \mr{4}{47.73}  \\
	IG4     & 84.83$\pm$2.96     & 27.09$\pm$8.55     & 88.33$\pm$2.58     &   					&      				\\ 
	IGFSS4  & 84.56$\pm$2.97     & 24.67$\pm$6.96     & 90.67$\pm$2.89     &   					&      				\\
	VGFSS4  & 85.00$\pm$3.12     & 17.87$\pm$5.06     & 89.59$\pm$1.69     &   					&      				\\ \midrule
	Chi3    & 85.28$\pm$3.23     & 27.75$\pm$6.26     & 92.55$\pm$2.12     & \mr{4}{2385}  	& \mr{4}{73.87}  \\
	IG3     & 83.39$\pm$3.04     & 22.34$\pm$5.21     & 92.10$\pm$2.32     &   					&      				\\
	IGFSS3  & 84.56$\pm$2.53     & 22.53$\pm$4.90     & 92.37$\pm$2.84     &   					&      				\\
	VGFSS3  & 86.00$\pm$3.12     & 23.61$\pm$5.82     & 92.01$\pm$1.53     &   					&      				\\ \midrule
	Chi2    & 86.35$\pm$2.72     & 43.91$\pm$3.92     & 93.81$\pm$2.90     & \mr{4}{1192}  	& \mr{4}{86.94}  \\ 
	IG2     & 84.46$\pm$3.05     & 37.45$\pm$7.74     & 93.27$\pm$1.85     &   					&     				\\
	IGFSS2  & 85.18$\pm$2.89     & 42.18$\pm$6.41     & 93.81$\pm$2.02     &   					&     				\\
	VGFSS2  & 85.55$\pm$2.10     & 45.52$\pm$3.83     & \bf{95.07$\pm$1.55}&   					&     				\\ \midrule
	Chi1    & 71.18$\pm$4.34     & 73.15$\pm$4.94     & 74.68$\pm$4.75     & \mr{6}{7}     	& \mr{6}{99.92}  \\
	IG1     & 75.30$\pm$6.61     & 77.19$\pm$7.14     & 78.80$\pm$7.09     &      				&      				\\
	IGFSS1  & 87.07$\pm$2.96     & 86.98$\pm$3.07     & 88.51$\pm$2.51     &      				&      				\\
	VGFSS1  & 87.07$\pm$2.96     & 86.98$\pm$3.07     & 88.51$\pm$2.51     &      				&      				\\
	DCD. E. & \bf{93.63$\pm$2.48}& \bf{93.27$\pm$2.49}& 92.46$\pm$2.55     & 					& 					\\
	DCD. C. & 89.22$\pm$2.49     & 90.21$\pm$1.31     & 91.20$\pm$2.48     &      				&      				\\
	\bottomrule
	\end{tabular}
	}
\end{table}
\begin{table}[tb]
  \caption{Experiments results with Cluster Dataset}
  \label{tab:results_cluster}
  \centering
  \resizebox{\columnwidth}{!}{
	\begin{tabular}{lccccc}
	\toprule
	\textbf{Tech}				 & \textbf{SVM}			& \textbf{KNN} 	& \textbf{RF} 	& \textbf{Feat.} 	& \textbf{Red(\%)} \\
	\midrule
	Orig    & 71.30$\pm$4.53    & 10.15$\pm$2.56    & 60.54$\pm$3.88     & 7630  			& 34.67      		\\ \midrule
	Chi4    & 69.85$\pm$4.59    & 36.13$\pm$3.74    & 61.02$\pm$4.52     & \mr{4}{6104} 	& \mr{4}{47.73}  \\
	IG4     & 69.61$\pm$3.88    & 35.95$\pm$3.05    & 61.02$\pm$5.68     &   				&       			\\ 
	IGFSS4  & 69.61$\pm$4.32    & 12.69$\pm$3.82    & 61.21$\pm$3.21     &   				&       			\\
	VGFSS4  & 69.67$\pm$3.56    & 15.53$\pm$5.71    & 62.54$\pm$4.40     &  				&       			\\ \midrule
	Chi3    & 70.64$\pm$4.48    & 31.48$\pm$3.71    & 66.58$\pm$3.40     & \mr{4}{3052}  & \mr{4}{73.87}  \\
	IG3     & 69.13$\pm$3.96    & 16.18$\pm$5.92    & 65.56$\pm$3.37     &  				&       			\\
	IGFSS3  & 70.64$\pm$3.66    & 29.91$\pm$7.29    & 66.83$\pm$3.69     &  				&       			\\
	VGFSS3  & 69.31$\pm$4.32    & 25.87$\pm$5.16    & 66.77$\pm$4.51     &  				&       			\\ \midrule
	Chi2    & 71.54$\pm$2.88    & 48.34$\pm$3.82    & 70.27$\pm$3.69     & \mr{4}{1526}  & \mr{4}{86.94}  \\ 
	IG2     & 70.40$\pm$3.57    & 32.57$\pm$4.11    & 70.09$\pm$3.66     &   				&       			\\
	IGFSS2  & 72.81$\pm$4.20    & 43.56$\pm$6.32    & 71.90$\pm$3.58     &   		 		&       			\\
	VGFSS2  & 71.24$\pm$4.58    & 33.72$\pm$4.96    & 70.57$\pm$3.35     &   		 		&       			\\ \midrule
	Chi1    & 55.65$\pm$2.96    & 55.71$\pm$3.43    & 57.40$\pm$3.04     & \mr{6}{7}  	& \mr{6}{99.92}  \\
	IG1     & 54.32$\pm$4.01    & 53.47$\pm$4.60    & 57.04$\pm$3.87     &      			   					\\ 
	IGFSS1  & 57.52$\pm$3.25    & 56.37$\pm$3.13    & 57.70$\pm$3.08     &      			       				\\
	VGFSS1  & 54.80$\pm$4.07    & 53.35$\pm$4.32    & 56.19$\pm$4.15     &      			      				\\
	DCD. E. & \bf{78.91$\pm$2.7}& \bf{77.4$\pm$2.81}& 77.46$\pm$3.06     & 					 					\\
	DCD. C. & 73.77$\pm$2.55    & 74.32$\pm$2.95    & \bf{78.49$\pm$2.46}&      			      				\\
	\bottomrule
	\end{tabular}
	}
\end{table}

The results obtained with our experiments are summarized on Tables \ref{tab:results_reuter}-\ref{tab:results_cluster}. The \emph{Red.} column represents the percentage of feature reduction obtained with the number of features described on column \emph{Feat.} compared with the total number of features before pre-processing. The best results are marked in bold. 

The first thing to notice from these tables is that DCD.C obtained overall better results than DCD.E for the Reuters and Newsgroups (Tables \ref{tab:results_reuter} and \ref{tab:results_news}) datasets. This is the expected behavior since the representative vectors are created as a sum of other vectors which implies that the intention is to capture a given direction close to most documents of the corresponding class. As such, the Cosine similarity is more compatible with this intuition.

On the other hand, for the Genes and Clusters datasets (Tables \ref{tab:results_genes} and \ref{tab:results_cluster}), the Euclidean distance performed better than the Cosine similarity, the reason why should be investigated.

Comparing to the other approaches, the DCD obtained the best results with the only exception of the combination of Random Forest with $20\%$ of the features selected by VGFSS. Despite that, both DCD versions achieved a very close result but with a much higher dimensionality reduction.

Finally, we can see that the application of DCD renders a much better result than the use of the original features set, thus indicating that the transformed features capture the information regarding the different classes.

In Figures \ref{fig:graphs_DCD_SVM}-\ref{fig:graphs_DCD_KNN} we can see the compromise between accuracy and reduction achieved by each combination of algorithm, parameters, and dataset. These figures highlight the interesting property of DCD of maximizing both the reduction and accuracy for the classification task.

Tables \ref{tab:words_reuter}-\ref{tab:words_cluster} give a detailed analysis of each representative vectors from all datasets used. In this tables were gathered the top 10 words with the highest intensity of each class. Observing Table \ref{tab:words_reuter}, referring to the dataset News, we noticed that many words could characterize the class that the representative vector expresses. For example, some of the words that have the highest intensity in the sci space class are space, orbit, mission, planet, earth, moon. Notice that these words have a certain relation to the class of their representative vector. We can observe the same pattern when we look at Table \ref{tab:words_reuter} which has the top 10 words of  Reuter dataset. This analysis is critical because it shows that representative vectors can capture the words that have the greatest and least contribution within a class, allowing to explore several unique characteristics of each class.

The relationship of the most significant words with class names representation is not very clear in Genes and Cluster datasets (Tables \ref{tab:words_genes} and \ref{tab:words_cluster}) because the class names do not express common words of the language. On the other hand, we can use the representative vectors to try to extract a name that best defines that class.

\begin{table}[htb]
  \caption{Top words per class on Reuter dataset}
  \label{tab:words_reuter}
  \centering
  \resizebox{\columnwidth}{!}{
  \begin{tabular}{ll}
    \toprule
    \textbf{Class label} & \mc{1}{c}{\textbf{Top 10 words}}       													  \\
    \midrule  
    ship	   	& gulf, attack, iran, ship, iranian, platform, oil, tanker, port, kuwait 		    \\
    corn	   	& nil, tonn, corn, wk, prev, export, import, maiz, soybean, wheat				        \\
    money-fx	& dollar, rate, bank, currenc, market, louvr, baker, dealer, bundesbank, accord \\
    interest	& rate, bank, pct, market, monei, bundesbank, interest, fed, cut, dai 			    \\
    acq	    	& share, compani, offer, acquir, stock, pct, group, dlr, acquisit, common 		  \\
    earn	    & mln, ct, net, loss, shr, profit, billion, dlr, rev, qtr 						          \\
    grain		  & nil, tonn, wheat, grain, export, crop, import, corn, wk, prev 				        \\
    trade		  & trade, japan, ec, deficit, export, import, billion, surplu, tariff, japanes 	\\
    wheat		  & tonn, wheat, export, crop, grain, season, soybean, mln, usda, stock 			    \\
    crude		  & oil, opec, crude, barrel, iran, price, gulf, bpd, attack, iranian 			      \\
  \bottomrule
  \end{tabular}
  }
\end{table}

\begin{table}[htb]
  \caption{Top words per class on 20NewsGroup dataset}
  \label{tab:words_news}
  \centering
  \resizebox{\columnwidth}{!}{
  \begin{tabular}{L{2.5cm}l}
   \toprule
    \textbf{Class label} & \mc{1}{c}{\textbf{Top 10 words}}                                   \\
    \midrule  
    sci crypt                   & kei, clipper, chip, encryption, govern, escrow, secur, wiretap, algorithm, nsa    \\
    sci med                     & patient, medic, cancer, vitamin, diseas, infect, hiv, doctor, drug, treatment     \\
    rec sport baseball          & game, player, hit, basebal, pitch, team, pitcher, clutch, gant, bat               \\
    alt atheism                 & moral, god, atheist, theism, belief, atheism, religion, islam, christian, livesei \\
    comp windows x              & do, window, server, openwindow, xterm, xview, file, font, widget, xv              \\
    sci space                   & space, orbit, mission, planet, shuttl, hst, earth, moon, henri, nasa              \\
    soc religion christian      & god, homosexu, christian, sin, church, christ, jesu, love, paul, bibl             \\
    comp os ms-windows misc     & window, file, do, os, microsoft, ms, driver, run, mous, font                      \\
    misc forsale                & sale, comic, wolverin, ship, offer, drive, price, cd, forsal, art                 \\
    comp graphics               & imag, graphic, jpeg, file, format, bit, gif, ftp, program, pub                    \\
    rec motorcycles             & bike, dod, ride, dog, bmw, motorcycl, rider, sun, shaft, east                     \\
    sci electronics             & batteri, copi, circuit, led, concret, board, acid, tape, protect, disk            \\
    talk politics.guns          & gun, fbi, fire, stratu, batf, atf, koresh, govern, waco, bd                       \\
    comp sys ibm pc hardware    & scsi, drive, dx, card, modem, id, cpu, bu, pc, mhz                                \\
    talk religion misc          & jehovah, god, elohim, christian, lord, jesu, mcconki, bibl, christ, sandvik       \\
    rec autos                   & car, engin, ford, mustang, diesel, clutch, dealer, geico, oil, auto               \\
    talk politics mideast       & armenian, muslim, israel, arab, jew, turkish, jewish, adl, isra, palestinian      \\
    rec sport hockey            & game, team, hockei, pit, espn, det, fan, playoff, goal, plai                      \\
    comp sys mac hardware       & mac, monitor, appl, mhz, drive, disk, system, problem, centri, lc                 \\
    talk politics misc          & myer, presid, cramer, stephanopoulo, homosexu, peopl, gai, ms, optilink, clayton  \\
  \bottomrule
  \end{tabular}
  }
\end{table}
\begin{table}[htb]
  \caption{Top words per class on Genes dataset}
  \label{tab:words_genes}
  \centering
  \resizebox{\columnwidth}{!}{
  \begin{tabular}{ll}
    \toprule
    \textbf{Class label} & \mc{1}{c}{\textbf{Top 10 words}}                                   \\
    \midrule  
    041			    &	pho, phosphatas, promot, nucleosom, acid, phosphat, secret, chromatin, activ, express 	\\
	  612			    &	telomer, rap, silenc, rapp, bind, sir, sirp, dna, domain, site 							            \\
	  344			    &	suc, invertas, snf, glucos, repress, ssn, secret, hxk, sucros, gene 					          \\
	  388			    &	cytochrome, cyc, iso, heme, hap, variant, denatur, structur, state, cyp 				        \\
	  211			    &	atpas, membran, pma, plasma, atp, ph, enzyme, proton, mutant, pump 						          \\
	  111			    &	ho, mat, dsb, mate, alpha, switch, sin, endonucleas, recombin, mata 					          \\
	  185			    &	rad, recombin, repair, strand, reca, homolog, dna, human, meiotic, ssdna 				        \\
   \bottomrule
  \end{tabular}
  }
\end{table}
\begin{table}[htb]
  \caption{Top words per class on Cluster dataset}
  \label{tab:words_cluster}
  \centering
  \resizebox{\columnwidth}{!}{
  \begin{tabular}{ll}
  \toprule
    \textbf{Class label} & \mc{1}{c}{\textbf{Top 10 words}}                                   \\
    \midrule  
   mcm	 		  &	pho, mcm, glucos, gpa, signal, phosphatas, pheromon, protein, acid, transcript 	\\
	 sic1	 	    &	chitin, rme, im, rmep, ch, invas, sporul, sic, ty, tec 							            \\
	 histone		&	ho, dsb, mat, switch, recombin, histon, break, mate, strand, endonucleas 		    \\
	 cln2	 	    &	rad, dna, polymerase, telomer, replic, repair, pol, cdc, telomeras, checkpoint 	\\
	 clb2	 	    &	cdc, myosin, apc, plk, kinas, chitin, spindl, bud, cyclin, polo 				        \\
	 mat	 		  &	alpha, factor, receptor, pheromon, rg, sst, agglutinin, ste, mf, mate 			    \\
	 met	 		  &	met, methionin, sulfurylas, oah, enzyme, sulfur, gsh, ap, atp, cysteine 		    \\
  \bottomrule
  \end{tabular}
  }
\end{table}

\section{Conclusion and Future Work}\label{sec:conclusion_and_future_work}

In this paper, we have proposed a new feature extraction algorithm, called DCDistance, that reduces the number of features to the number of distinct class labels of the dataset.

This algorithm simply creates representative vectors as the sum of training vectors grouped by each class and then uses these vectors to create a distance vector with each element being the distance of a document to a representative vector.

Since the features set is reduced to the number of class labels, this algorithm creates a reduction of more than $99\%$ of the original feature set. With a proper choice of data structures, this algorithm can be implemented with a linear complexity w.r.t. the number of documents.

The experimental results compared DCDistance against four other techniques and, also, the use of the original set of features. The results showed that the proposed algorithm outperforms each of the contenders by maximizing both the reduction of the features set and the accuracy of the machine learning model.

By observing the words corresponding to the features with the highest values in the representative vectors, we perceived a high correlation with the topics pertaining to the class. This property allows exploring other applications for these attributes, such as summarization or model interpretability.

For future research, we will explore other forms of aggregation to create the representative vectors and different distance metrics. And, as mentioned, we will also explore the interpretability induced by these representative vectors.

\section*{Acknowledgments}

\bibliographystyle{model2-names}
\bibliography{refs}



\end{document}